\documentclass[aps,prd,twocolumn,superscriptaddress,nofootinbib,eqsecnum]{revtex4}

\pdfoutput=1

\usepackage{amsfonts}
\usepackage{amsmath}
\usepackage{amssymb}
\usepackage{graphicx,color}
\usepackage{float}
\usepackage{hyperref}
\usepackage{subfigure}
\usepackage{dcolumn}

\begin{document}

\title{Warm inflation as a way out of the swampland}

\author{Meysam Motaharfar}
\email{mmotaharfar2000@gmail.com}
\affiliation{Department of Physics, Shahid Beheshti University, G. C., Evin,Tehran 19839, Iran}

\author{Vahid  Kamali}
\email{vkamali@basu.ac.ir}
\affiliation{Department of Physics, Bu-Ali Sina University, Hamedan  65178, 016016, Iran}

\author{Rudnei O. Ramos}
\email{rudnei@uerj.br}
\affiliation{Departamento de Fisica Teorica, Universidade do Estado do Rio de Janeiro, 
20550-013 Rio de Janeiro, RJ, Brazil }

\begin{abstract}

We discuss how dissipative effects and the presence of a thermal
radiation bath, which are inherent characteristics of the warm
inflation dynamics, can evade the recently proposed Swampland
conjectures.  Different forms of dissipation terms, motivated by both
microphysical quantum field theory and phenomenological models, are
discussed and their viability to overcome the assumed Swampland
constraints is analyzed.

\end{abstract}

\maketitle

\section{Introduction} 

The cosmological observational data unanimously confirm that the
Universe is expanding, spatially flat, homogeneous, and isotropic on
large scales and the Large Scale Structure (LSS) originated from
the inhomogeneities of primordial origin, which are adiabatic and Gaussian and with a quasi-invariant scale power
spectrum~\cite{Aghanim:2018eyx}. The current paradigm for the
evolution of the early Universe, cosmological
inflation~\cite{inflation}, consisting of a short period of quasi--de
Sitter accelerated expansion phase,  is able to successfully explain
several puzzles of the standard big bang cosmology. Inflation solves
the horizon and flatness problems and at the same time accounts for the
origin of the inhomogeneities in the Universe able to lead to the LSS
formation, based on causal physics.  Although inflation generally
occurs when the inflaton field is super-Planckian, especially when
considering monomial chaotic potentials, when the currently
observable scale exits the Hubble horizon, the energy scale is much
smaller than the Planck scale. Therefore, it is assumed that inflation
can be well described by some low-energy effective theory
(EFT). However, this does not mean that any inflationary model, in the
language of the effective theory, can be ultraviolet complete. In
fact, consistently embedding the EFT of gravity into a quantum theory, in
particular in the context of string theory, requires distinguishing
consistent low-energy EFT coupled to gravity from inconsistent
counterparts. The latter are deemed to inhabit the so-called {\it
  Swampland}~\cite{Vafa, Ooguri:2006in}. Hence, having correct
criteria to identify the boundary between the landscape and the
Swampland has resulted in a series of conjectures known as weak
gravity~\cite{Palti:2017elp} and Swampland
conjectures~\cite{Obied:2018sgi}, motivated by black hole
physics~\cite{Arkani-Hamed} and string
compactification~\cite{Ooguri:qwFreivoge:swBrennan:sd}. Recently, it
was proposed that an effective field theory to be consistently
embedded in quantum gravity must satisfy these conjectures. These
conjectures then turn out to lead to constraints in the allowed range
of field excursion and the potential field gradient. In turn, these
conjectures have lead to severe constraints on the possible inflation
models able to be described in terms of effective models of
inflation~\cite{Agrawal:2018own}.  The two Swampland conjectures are
as follows.

$\bullet$ {\it The Swampland distance conjecture (SDC)}.---The scalar
field excursions in field space are bounded from
above~\cite{Ooguri:2006in}:
\begin{align}
\frac{|\Delta\phi|}{M_{ \rm Pl}} < \Delta \sim {\cal O}(1),
\label{SDC}
\end{align}
where $\Delta$ is some constant of the order of unity, that depends on the
details of the string model compactification.  $M_{ \rm Pl} \equiv
(8\pi G)^{-1/2} \simeq 2.4 \times 10^{18}$ GeV is the reduced
Planck mass.

$\bullet$ {\it The Swampland de Sitter conjecture (SdSC)}.--- The field
gradient of the potential $V$ (with $V>0$) in any consistent theory of
gravity should satisfy the lower bound~\cite{Obied:2018sgi}:
\begin{align}
M_{ \rm Pl}\left(\frac{|\nabla_{\phi} V|}{V}\right)>
c\sim\mathcal{O}(1),
\label{SdSC}
\end{align}
with $c$ being another constant also of the order of unity. 

It should be noticed that Eq.~(\ref{SdSC}) not only requires steep
potentials but also forbids models including extrema or plateaus, for
which $\nabla_\phi V/V\rightarrow 0$ in one or more points in field
space. Although the authors in Ref.~\cite{Dvali:dc} have shown that
there is still controversy on the specific value for $c$, which may
make the discussion irrelevant, we will assume Eq.~(\ref{SdSC}) for
the sake of argument. 

Hence, by taking at face value and by assuming the validity of the
above  two conjectures, they may have serious cosmological
implications as discussed by many authors
recently~\cite{Agrawal:2018own,Andriot:2018eptRoupec:2018mbnAndriot:2018wzkColgain:2018wgkHeisenberg:2018yae},
in particular, for inflationary
physics~\cite{Garg:2018reuBrown:2015ihaMatsui:2018bsyBen-Dayan:2018mheKinney:2018nny,
  Dias:2018ngv}.  In fact, possible inflation theories, especially
slow-roll single field inflation, are tightly constrained. Thus, more
complicated inflation models, or nonstandard physics of the early
Universe, such as chromonatural inflation~\cite{Agrawal:2018},
multifield inflation~\cite{Achucarro:2018vey}, or curvaton
scenarios~\cite{Kehagias:2018uem} seem to be preferred by these two
conjectures.  One should notice that the validity of the above
Swampland conjectures has also been questioned
recently~\cite{Akrami:2018ylq,Conlon:2018eyr}.  Independent of the
real applicability or validity of these conjectures, we believe it is
still worth seeking which inflationary model realizations or pictures
might overcome them. This, in particular, might teach us something about
the possible realization of these models in the context of effective
field theories and a possible connection of them with more fundamental
theories, like string theory.  In view of these motivations, in this
work, by assuming the validity of the above two Swampland conjectures,
we would like to discuss them in the context of the warm inflation
picture~\cite{Berera:1995ie} as a possible way of realizing in  a
satisfactory way the conditions given by Eqs.~(\ref{SDC}) and
(\ref{SdSC}). As we are going to show, by accounting for the
nonequilibrium dissipative processes emerging from the microscopic
dynamics of the inflaton field during warm inflation, which can
produce a quasiequilibrium  thermal radiation bath during inflation
(for reviews on warm inflation, see, e.g.,
Refs.~\cite{Berera:2008ar,BasteroGil:2009ec}), the  dynamics of
inflation can change considerably at both the background and
perturbation levels compared with the case where these effects are
neglected.  Consequently, warm inflation can lead naturally to
different consistency relations compared to the case when these
dissipative effects are negligible, like in the usual cold inflation
scenario. Thus, warm inflation can possibly lead to conditions able, in
principle, to satisfy Eqs.~(\ref{SDC}) and (\ref{SdSC}). 
  This ability for warm inflation possibly evading the Swampland
  conditions was  first noticed in Ref.~\cite{Das:2018hqy}.  In this
work we would like to establish whether warm inflation is indeed able
to satisfy Eqs.~(\ref{SDC}) and (\ref{SdSC}) and under which
conditions and for which possible models this can indeed happen.

This paper is organized as follows. In Sec.~\ref{sec2}, we briefly
review the warm inflation dynamics and its relevant equations, which
we will be using in this work. Next, in Sec.~\ref{sec3}, we will then
study different dissipation mechanisms that can lead to the warm
inflation dynamics, either directly motivated from particle physics
realizations or phenomenological motivated ones.  We will then discuss
these models in the context of the Swampland conjectures. {}Finally,
in Sec.~\ref{concl}, we give our conclusions and final comments.

\section{Warm inflation dynamics}
\label{sec2}

The dynamics of inflation can go through two different pictures: {\it
  cold inflation} (CI) and {\it warm inflation} (WI), depending on
whether nonequilibrium  dissipative processes due to the couplings of
the inflaton field with other field degrees of freedom are negligible
or not during inflation. In fact, dissipation processes determine how
ultimately the vacuum energy density, stored in the inflaton field, ends
up converting into radiation, thus allowing the Universe to  make a
transition from the accelerating phase to the radiation-dominated
epoch. The first and more conventional picture for inflation is the
isentropic CI, where the inflaton field can be treated as essentially
isolated from interacting with other subdominant quantum fields,
whereby any previous preinflationary radiation energy density
drastically decreases. Later, at the end of inflation, the inflaton
starts oscillating around the minimum of its potential and
progressively dissipates its energy into other relativistic light
degrees of freedom that thermalizes and provides the radiation bath
required by the standard big bang cosmology. Thus, CI necessarily
requires a phase of (p)reheating (for some recent reviews on the
(p)reheating theory, see, e.g., Refs.~\cite{ABCM,Karouby}, and
references therein).  In contrast to this picture,  in WI, the inflaton
interactions with other subleading quantum fields can be strong
enough in such a way that their effects may not become
negligible. Therefore, the state of the inflationary universe can
become one not that of a quasiperfect vacuum state like in CI, but
rather an excited statistical state, with a thermal state being the
most examined~\cite{Berera:2008ar,BasteroGil:2009ec}.  Even so, in WI
the vacuum energy is still the dominant component for the accelerated
expansion to take place. Consequently, non-negligible dissipative
processes can take place not only after but also during the slow-roll
phase of inflation, whereby a quasiequilibrium thermal radiation bath
is concurrently generated throughout inflation. This state can be such
to compensate the supercooling phase observed in CI and radiation can
smoothly be produced and it becomes the dominant energy component of
the Universe at the end of the inflationary expansion. 

The total energy density in WI is then given by 
\begin{equation}
\rho = \frac{{\dot \phi}^2}{2} + V(\phi) + \rho_R,
\label{rho}
\end{equation}
which accounts for the presence of the radiation fluid, with energy
density $\rho_R$, and the scalar field (the inflaton) $\phi$, with
some potential $V(\phi)$.  The inflaton field $\phi$ and the radiation
energy density $\rho_R$ form a coupled system in WI dynamics, with
background evolution equations given, respectively, by
\begin{eqnarray}
&& \ddot \phi + 3 H \dot \phi + \Upsilon(\phi, T) \dot \phi +
  V_{,\phi}=0,
\label{eqphi}
\\ && \dot \rho_R + 4 H \rho_R = \Upsilon(\phi, T) \dot \phi^2,
\label{eqrhoR}
\end{eqnarray}
where $\Upsilon(\phi, T)$ is the dissipation coefficient in WI, which
can be a function of the temperature and/or the background inflaton
field. We will define below explicitly the dissipation coefficient
forms we will be considering in this work.  {}For a radiation bath of
relativistic particles, the radiation energy density is given by
$\rho_R=\pi^2 g_* T^4/30$, where $g_*$ is the effective number of
light degrees of freedom  ($g_*$ is fixed according to  the
dissipation regime and interactions form used in WI).  In WI, it is
usual to define the ratio of the dissipation coefficient with the
Hubble rate $H$ as
\begin{equation}
Q= \frac{\Upsilon(T,\phi)}{3 H},
\label{Q}
\end{equation}
which in some sense parametrizes the strength of the WI dissipative
processes.  The dissipation coefficient $\Upsilon$ in
Eqs.~(\ref{eqphi}) and (\ref{eqrhoR})  embodies the microscopic
physics resulting from the interactions between the inflaton and  the
other fields that can be present and accounts for the nonequilibrium
dissipative processes arising from these interactions. The dissipation
coefficient $\Upsilon$ is, in general, well defined in the context of
nonequilibrium quantum field theory
methods~\cite{Berera:2008ar,BasteroGil:2010pb,BasteroGil:2012cm}.  The
effect of the dissipation term in the above equations can be such as
to be able to sustain a temperature $T\ge H$ throughout inflation  for
$\dot\phi\gg H^{2}$, even for $\Upsilon \ll H$ (i.e., $Q \ll 1$),
without violating the slow-roll conditions.  In particular, the
additional friction term in Eqs.~(\ref{eqphi}) and (\ref{eqrhoR})
modifies the usual slow-roll conditions to $\{\epsilon_{V},
\eta_{V}\}\ll 1+Q$, where $\epsilon_{V} = M^{2}_{\rm
  Pl}(V_{,\phi}/V)^{2}/2$ and $\eta_{V} = M^{2}_{\rm Pl}
V_{,\phi\phi}/V$ are the usual (inflaton potential) slow-roll
parameters. Therefore, the conventional Hubble slow-roll parameter
$\epsilon_{H}$ changes to
\begin{align}
\epsilon_{H} = -\frac{\dot H}{H^{2}} \simeq \frac{\epsilon_{V}}{1+Q}.
\label{epsilonH}
\end{align}
Thus, the accelerated inflationary dynamics terminates when
$\epsilon_{H}=1$ or, equivalently, when $\epsilon_{V} = 1+Q$.
Moreover, the ratio of radiation to inflaton energy density in the
slow-roll regime is roughly given  by
\begin{align}
\frac{\rho_{R}}{\rho_{\phi}} \simeq \frac{1}{2}
\frac{\epsilon_{V}}{1+Q} \frac{Q}{1+Q}.
\end{align}
Therefore, during inflation, the energy density associated with the
inflaton field still dominates over the one from the radiation,
$\rho_{\phi}\gg \rho_{R}$. Hence, radiation is somewhat still
suppressed, as we should expect. Even though this condition might be
satisfied for weak dissipation ($Q\ll 1$), at the end of inflation for
$\epsilon_{V} \sim {1+Q}$ and ${\rho_{R}}/{\rho_{\phi}} \simeq
({{1+Q}})Q/2$, radiation can become the dominant component.
Consequently, the Universe can smoothly enter into a radiation-dominated era without the need of a reheating phase $a \ priori$. In fact,
steep potentials without extrema and plateau can also be embedded in
WI, as the Swampland criteria typically prefer.

\subsection{The Swampland and a lower bound on the value of the dissipation}

Slow-roll CI models are in tension with the SdSC, since in such models
$\epsilon_{H} \simeq \epsilon_{V}$ and Eq.~(\ref{SdSC}) would imply
that $\epsilon_V > c^2/2$, thus preventing an accelerating
expansion. Even though the condition $c > 1$ can, in principle, be
relaxed~\cite{Kehagias:2018uem}, one notices that WI can break the
approximate relation between $\epsilon_H$ and $\epsilon_V$ as
discussed above, from Eq.~(\ref{epsilonH}), allowing  accelerated
expansion even if $\epsilon_V > 1$, provided that $Q >1$.  One notes
also the SdSC condition, Eq.~(\ref{SdSC}), targets the heart of CI
models based on the requirement of the standard (p)reheating
mechanism, which usually occurs around the minimum of the potential,
$V_{,\phi} = 0$. As already discussed above, the continuous radiation
production in WI due to the dissipative effects can naturally replace
the usual reheating mechanism by smoothly ending inflation already in
a radiation-dominate regime. Thus, WI does not require, in general, a
potential with a minimum or that $V_{,\phi} = 0$ as a condition for
reheating.  {}Furthermore, note that we can also express the SDC
condition, Eq.~(\ref{SDC}), for the field excursion in terms of the
slow-roll parameter  $\epsilon_V$ and the  number of $e$-folds $N$ as
(using $dN = H dt$)
\begin{align}\label{A3}
\frac{\Delta\phi}{M_{ \rm Pl}} = \frac{\dot\phi}{H}  N \simeq
\frac{\sqrt{2\epsilon_{V}}}{(1+Q)} N.
\end{align}
Note that in CI ($Q\equiv 0$) we can have plateau-type potentials
for which $\epsilon_{V} \ll 1$ and $N \gg 60$, which is more than
necessary for solving the usual big bang cosmological problems,
leading to very small field excursions and thus satisfying the
condition Eq.~(\ref{SDC}). However, plateau-type potentials are
strongly disfavored by the second Swampland condition
Eq.~(\ref{SdSC}). Even so, we see that the change of the slow-roll
parameters by the dissipation ratio $Q$ in Eq.~(\ref{A3}) offers
again a way out of the SdSC condition if $Q \gg 1$ and $N$ is still
within the required number of $e$-folds, $N\sim 60$.   Thus, putting
these conditions together in the context of WI, such as to
simultaneously satisfy both conjectures, we are lead to a  lower
bound on the dissipation ratio $Q$ as given by
\begin{align}\label{A1}
Q>\frac{c}{\Delta}  N-1. 
\end{align}
We can regard Eq.~(\ref{A1}) as a general lower bound on
  $Q$ for  WI models, regardless of the exact value of $c$ and
  $\Delta$,  to be able to simultaneously satisfy both Swampland
  conjectures.  Although there is still an uncertainty on the exact
  values for $c$ and $\Delta$, the above bound shows that WI is on the
  landscape, given an appropriate value for the dissipation ratio $Q$,
  and even for $c$ and $\Delta$ taken to be of the order of unity and with $N
  \sim 60$. By relaxing these parameters, we can also make it easier for
  warm inflation to satisfy the Swampland conjectures for smaller
  values of $Q$, as we will show  in Sec.~\ref{sec3} below.

\subsection{The Swampland, WI, and the scalar and tensor perturbations}

In WI, not only the background dynamics gets modified by the
dissipative effects, but also the perturbations get modified due to
the presence of dissipation and the radiation bath. Especially, the
primordial power  spectrum can be strongly influenced by these effects
(see, e.g.,
Refs.~\cite{Hall:2003zp,Graham:2009bf,BasteroGil:2011xd,Bastero-Gil:2014jsa,Bastero-Gil:2014raa,Visinelli:2014qla}).
{}For instance, the primordial power spectrum for WI at horizon
crossing can be expressed in the form (see, e.g.,
Ref.~\cite{Benetti:2016jhf} and references therein)
\begin{equation} \label{Pk}
\Delta_{{\cal R}}(k/k_*) =  P_0(k/k_*) {\cal F} (k/k_*),
\end{equation}
where we indicate with a subindex ``$*$" those quantities evaluated at
the horizon crossing.  The contribution $P_0(k/k_*)$ in Eq.~(\ref{Pk}) is
the standard CI result:
\begin{equation}
P_0(k/k_*) \equiv  \left(\frac{ H_{*}^2}{2 \pi\dot{\phi}_*}\right)^2 ,
\end{equation}
while ${\cal F} (k/k_*)$ in Eq.~(\ref{Pk})  is the WI modification to
the primordial power spectrum:
\begin{equation}
{\cal F} (k/k_*) \equiv  \left(1+2n_* + \frac{2\sqrt{3}\pi
  Q_*}{\sqrt{3+4\pi Q_*}}{T_*\over H_*}\right) G(Q_*).
\label{calF}
\end{equation}
In the latter equation, $n_*$ denotes the inflaton statistical
distribution due to the presence of the radiation bath.  {}For a
thermal equilibrium distribution,  it assumes the Bose-Einstein
distribution form, $n_* =1/[\exp(H_*/T_*) -1]$.   The function
$G(Q_*)$ in Eq.~(\ref{calF}) accounts for the effect of the coupling
of the inflaton fluctuations with
radiation~\cite{Graham:2009bf,BasteroGil:2011xd,Bastero-Gil:2014jsa},
which we will specify in more detail in the next section below.  The
scalar spectral amplitude value at the pivot scale $k_*$ is set by the
cosmic microwave background (CMB) radiation data as  $\Delta_{{\cal
    R}}(k=k_*) \simeq 2.2 \times 10^{-9}$, with $k_*=0.05 {\rm
  Mpc}^{-1}$ as considered, e.g.,  by the Planck
Collaboration~\cite{Aghanim:2018eyx}.

While the primordial scalar curvature perturbation in WI gets modified
according to Eq.~(\ref{Pk}), the tensor perturbations spectrum is
simply of the CI form, $\Delta_{T} = 2 H_*^2/(\pi^2 M_p^2)$.  This is
so because of the weakness of the gravitational interactions. Hence,
the spectrum of primordial gravitational waves remains unaffected by
the dissipative dynamics of WI~\cite{Ramos:2013nsa} (see also
Ref.~\cite{Li:2018wno} for a recent study on the possible changes of
the tensor spectrum in WI).  The tensor-to-scalar ratio $r$,
\begin{equation}
r= \frac{\Delta_{T}}{\Delta_{{\cal R}}},
\label{eq:r}
\end{equation}
leads now to a modified consistency relation in WI when compared to
the CI result,
\begin{align}\label{A2}
r= \frac{16\epsilon_{V}}{(1+Q_{\star})^{2}} {\cal F}^{-1} (k/k_*).
\end{align}
Combining this result with Eq.~(\ref{A3}) and the two Swampland
conjectures Eqs.~(\ref{SDC}) and (\ref{SdSC}), we are lead to the
following requirement:
\begin{align}\label{Aq}
\frac{8c^{2}}{(1+Q)^{2}} \frac{1}{\mathcal{F}} <r<\frac{8\Delta^{2}}{
  N^{2}} \frac{1}{\mathcal{F}}.
\end{align}
Note that for $Q=0$ (when $\mathcal{F}\to 1$), then Eq.~(\ref{Aq})
reduces to the relation obtained for CI when considering the Swampland
conjectures~\cite{Dias:2018ngv}.   Although it is rather
  difficult for both constraints in Eq.~(\ref{Aq}) to be
  simultaneously satisfied in CI for $\{c, \Delta\} \sim
  \mathcal{O}(1)$,  it can be possible to be arranged for both of them
  to be satisfied in WI. This can happen, for instance, if the
  dissipation ratio is larger than the $e$-folding number as we have
  already stated in Eq.~(\ref{A1}). If we relax on the $c$ and
  $\Delta$ values, it can also be arranged for Eq.~(\ref{Aq}) to be
  satisfied even for smaller values of $Q$, as we will show in
  Sec.~\ref{sec3}.  Besides, it is important to also check what the
  effects of such given values of $Q$, able to satisfy Eq.~(\ref{Aq})
  as required by the Swampland conditions, might have on the
  observables and whether they would still  be allowable when
  confronting some specific models  with the observational data. This
  is what we now focus on below.

\section{Some explicit model realizations and observational consequences}
\label{sec3}

Dissipation processes during WI modify not only the homogeneous
evolution of the inflaton field but also its inhomogeneous
fluctuations. Therefore, the curvature power spectrum gets modified in
different ways. {}First, it directly sources inflaton fluctuations,
yielding a Langevin equation that generalizes Eq.~(\ref{eqphi}) for an
inhomogeneous
field~\cite{Hall:2003zp,Graham:2009bf,BasteroGil:2011xd,Bastero-Gil:2014jsa,Bastero-Gil:2014raa,Visinelli:2014qla}.
In fact, the source of primordial perturbations stems from thermal
fluctuation in the radiation bath, which is then transferred to the
inflaton field as adiabatic curvature perturbations. Hence, WI
predicts that the origin of perturbations is thermal rather than
quantum. Inflaton quanta may also be thermally excited during
inflation and  acquire a Bose-Einstein rather than a vacuum phase
space distribution. {}Finally, inflaton and radiation fluctuations are,
in general, coupled due to the possible explicit temperature dependence
in the dissipative coefficient function $\Upsilon(\phi,T)$.  This can
be understood once the perturbations for WI are explicitly written
down~\cite{Graham:2009bf}. {}For example, if we parametrize the
dissipation coefficient as 
\begin{equation}
\Upsilon(\phi,T)= C_\phi T^s/\phi^{s-1},
\label{Upsilons}
\end{equation}  
where $s$ is some real number, the perturbed form for
Eq.~(\ref{Upsilons})  becomes
\begin{equation}
\delta \Upsilon = \left[ \frac{s}{4} \frac{\delta \rho_R}{\rho_R} -
  (s-1) \frac{\delta \phi}{\phi} \right] \Upsilon,
\label{pert}
\end{equation}
which explicitly makes the inflaton perturbations $\delta \phi$ to be
coupled with those of the radiation bath $\delta \rho_R$. As shown in
Ref.~\cite{Graham:2009bf}, this can generate in the curvature power
spectrum a growing mode (if $s> 0$) or a decreasing mode (if $s<0$),
as the dissipation ratio $Q$ increases.  {}For instance, the resulting
dimensionless power spectrum in WI becomes Eq.~(\ref{Pk}), with ${\cal
  F}$ given by Eq.~(\ref{calF}), and  the function $G(Q_*)$ in that
equation takes into account this effect of the coupling between
inflaton and radiation perturbations.  The strongest effect of this
term on the observables manifests through the spectral tilt $n_s$:
\begin{equation}
n_s -1 = \lim_{k\to k_*}   \frac{d \ln \Delta_{{\cal R}}(k/k_*) }{d
  \ln(k/k_*) },
\label{eq:n}
\end{equation}
which can be driven quickly to bluer values ($n_s > 1$) at larger
values of $Q$ when $s>1$ or lead to a redder spectrum when $s<0$.

In our analysis below, we will consider five forms for the dissipation
coefficient $\Upsilon$ that have been considered with some frequency
before in the literature and which are motivated by the
parameterization of the form of Eq.~(\ref{Upsilons}), namely

,(a) a dissipation coefficient with a cubic dependence in the
temperature,
\begin{equation}
\Upsilon_{\rm cubic} = C_{\rm cubic} T^3/\phi^2;
\label{cubic}
\end{equation}
(b) a dissipation coefficient with a linear dependence in the
temperature,
\begin{equation}
\Upsilon_{\rm linear} = C_{\rm linear} T;
\label{linear}
\end{equation}
(c) a dissipation coefficient inversely proportional to the
temperature,
\begin{equation}
\Upsilon_{\rm inverse} = C_{\rm inverse} \phi^2/T;
\label{inverse}
\end{equation}
(d) a dissipation coefficient that is constant,
\begin{equation}
\Upsilon_{\rm constant} = C_{\rm constant} M_{\rm Pl};
\label{constant}
\end{equation}
and (e) a dissipation coefficient proportional to the Hubble parameter,
\begin{equation}
\Upsilon_{\rm H} = C_{\rm H} H.
\label{UpsH}
\end{equation}
A dissipation coefficient of the form of $\Upsilon_{\rm cubic}$,
Eq.~(\ref{cubic}), emerges naturally for models of WI when  the
inflaton is coupled to heavy intermediate fields, that are in turn
coupled to light radiation
fields~\cite{Berera:2008ar,BasteroGil:2010pb,BasteroGil:2012cm}.  This
is obtained in the so-called low-temperature regime for WI, in which
the inflaton couples only to the heavy intermediate fields, whose
masses are larger than the radiation temperature, $M \gg T$, and, thus,
the inflaton gets decoupled from the radiation fields.  The
dissipation coefficient of the form of $\Upsilon_{\rm linear}$,
Eq.~(\ref{linear}), has been explicitly constructed
recently~\cite{Bastero-Gil:2016qru}, and it is based on a construction
used in Higgs phenomenology beyond the standard model, which uses a
collective symmetry where the inflaton is a pseudo-Goldstone boson.
In this case, the inflaton is directly coupled to the radiation fields
and gets protection from large thermal corrections due to the
symmetries obeyed by the model.  In these models, the dissipative
coefficient is derived in the high-temperature regime, where the
fields coupled to the inflaton are light with respect to the ambient
temperature, $M < T$.  The form of the dissipation coefficient as
given by Eq.~(\ref{inverse}) originated in the first studies of warm
inflation based on nonequilibrium quantum field theory
methods~\cite{Berera:1998gx,Berera:1998px} and where it was also
derived in a high-temperature regime, $M < T$. Though this type of
model is more difficult to explicitly construct as compared to
the two previous cases, we include it here as an example of
a dissipation regime that can lead to a suppression of the power
spectrum at larger $Q$ values and, thus, to a redder spectrum as
compared to the cubic and linear in $T$ cases, which both tend to lead
to a bluer spectrum at large $Q$. This is particularly important when
setting limits on the dissipation ratio based on the spectral tilt
$n_s$.  As in WI we typically have that the temperature decreases as
the end of inflation approaches, we will always have that eventually
the high-temperature regime used to derive the dissipation coefficient
like of the form of Eq.~(\ref{inverse}), will eventually transit to
the low-temperature form of Eq.~(\ref{cubic}) (see, e.g.,
Ref.~\cite{BasteroGil:2012cm}).  This motivates us to study also a
constant dissipation coefficient like in Eq.~(\ref{constant}), which
might describe this intermediate regime between the high- and low-temperature ones and if this regime lasts long enough. Finally, we
also include a dissipation coefficient of the form of Eq.~(\ref{UpsH}),
since it leads strictly to a constant dissipation ratio $Q$. Such a
form leading to a constant dissipation ratio is particularly useful
for deriving analytical results in WI and has been employed by many
authors before exactly because of that (see also
Ref.~\cite{Zhang:2009geHerrera:2014mcaJawad:2017gwa} for other forms
of dissipation coefficients).  It can be seen as a sort of
phenomenological dissipation form. In fact, many recent works dealing
with dissipation effects in the recent Universe, like describing
possible energy exchanges between the dark sectors, have been using
this type of dependence (see, e.g., Ref.~\cite{Bolotin:2013jpa} and
references therein), and it is also typically used to model dissipation
effects in the form of effective viscosities as
well~\cite{Barbosa:2017ojt}.

Assuming for demonstrative purposes an inflaton potential given by a
chaotic quartic potential\footnote{Although  strictly the form for
  this potential would not conform to the Swampland criterion SdSC,
  here we are interested only in the inflationary patch of the
  potential in the WI scenario. As also already explained before,  we
  are not interested in describing (p)reheating here around the
  minimum of the potential, since in WI dynamics this is not required
  $a \ priori$.  Thus, the dynamics around the minimum of the potential is
  not a concern to us here.}, $V(\phi) = \lambda \phi^4/4$, with
coupling constant $\lambda$ fixed by the primordial scalar spectrum
CMB amplitude according to the dissipation model used and spectrum
given by Eq.~(\ref{Pk}).   {}For the temperature-dependent
dissipations, the function $G(Q_*)$ appearing in Eq.~(\ref{calF}) can
then be found to be~\cite{Benetti:2016jhf}
\begin{eqnarray} \label{Gcubic}
G_{\rm cubic}(Q_*)\simeq 1+ 4.981 Q_*^{1.946}+ 0.127 Q_*^{4.330},
\end{eqnarray} 
for the cubic dissipation coefficient  $\Upsilon_{\rm cubic}$ and
\begin{eqnarray} \label{Glinear}
G_{\rm linear}(Q_*)\simeq 1+ 0.335 Q_*^{1.364}+ 0.0185Q_*^{2.315},
\end{eqnarray} 
for the linear dissipation coefficient $\Upsilon_{\rm linear}$, while
for the inverse in the temperature dissipation coefficient,
$\Upsilon_{\rm inverse}$, a similar numerical fitting as used to
derive the previous two cases~\cite{BasteroGil:2011xd} leads to the
result
\begin{eqnarray} \label{Ginverse}
G_{\rm inverse}(Q_*)\simeq \frac{1+ 0.4 Q_*^{0.77}}{\left(1+ 0.15
  Q_*^{1.09}\right)^2}.
\end{eqnarray}
{}For the dissipation terms independent of temperature,
Eqs.~(\ref{constant}) and (\ref{UpsH}), we can set this function $G$
equal to one.\footnote{There is a weak temperature dependence in
  Eq.~(\ref{UpsH}) through the radiation energy density contribution
  to the Hubble parameter, but during inflation, as already discussed,
  the dominant energy density component is that of the inflaton, so $H
  \propto \sqrt{V(\phi)}$ and we can still set $G_{\rm H} \sim 1$ with
  good accuracy.}

\begin{center}
\begin{figure}[!htb]
\includegraphics[width=8.6cm]{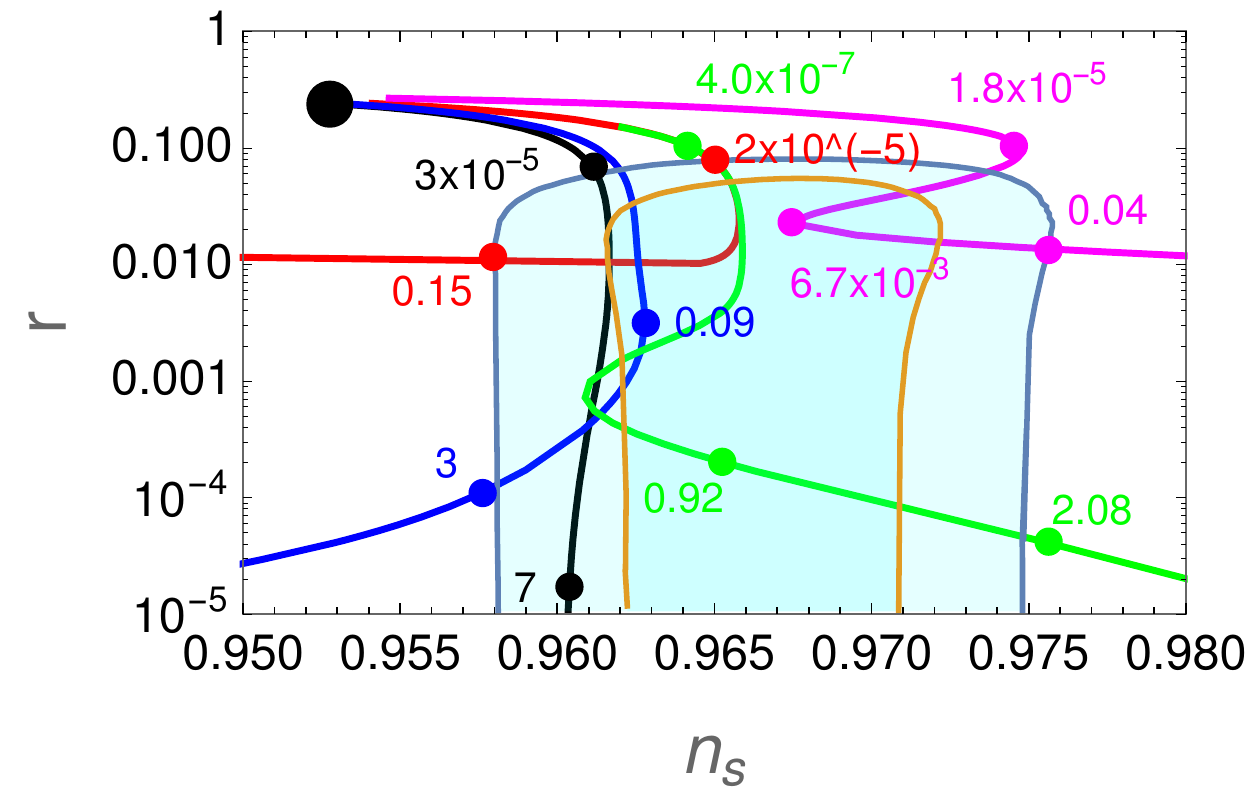}
\caption{The spectral index $n_s$ and the tensor-to-scalar ratio $r$
  in the plane $(n_s,r)$ for different values of the dissipation ratio
  $Q_*$ (indicated by the numbers next to the curves) and for the
  different dissipation forms considered in the text: the dissipation
  coefficient proportional to $H$ (black line), the constant
  dissipation coefficient (red line), the inverse in $T$ dissipation
  coefficient (blue line), the linear in $T$ dissipation coefficient
  (green line), and the cubic in $T$ dissipation coefficient (magenta
  line).  The large dot indicates the CI result ($Q=0$). The contours
  are for the $68\%$ and $95\%$ C.L. results from Planck 2018
  (TT+TE+EE+lowE+lensing+BK14+BAO data). }
\label{fignsXr}
\end{figure}
\end{center}

In {}Fig.~\ref{fignsXr}, we show the results obtained for both $r$ and
$n_s$ in WI for the five forms of dissipative coefficients  given by
Eqs.~(\ref{cubic}) - (\ref{UpsH}). {}For reference, for each
dissipation case, we have indicated some representative values of the
dissipation ratio $Q_*$. The datasets used for the contours are for
the $68\%$ and $95\%$ C.L. results from the recent Planck
2018~\cite{Akrami:2018odb}, and we have chosen the
TT+TE+EE+lowE+lensing+BK14+BAO data, which is the most restrictive
dataset.  {}For definiteness, we have also considered $N_*=60$ to
obtain the results shown in {}Fig.~\ref{fignsXr}.

Among the various dissipation cases studied, we see from
{}Fig.~\ref{fignsXr}  that the tensor-to-scalar ratio $r$ gets
naturally suppressed quickly as $Q$ increases for all forms of
dissipation terms considered, which is a natural feature of WI.
However, the main limiting effect for going to larger values of $Q$ is
on the spectral tilt.  {}Figure~\ref{fignsXr} shows that most of the
models  typically do not allow for large values of $Q$, $Q>1$, with
the maximum value reached for $Q_* \sim 2-3$, e.g., as in the cases
of the linear and inverse in $T$ dissipation coefficients, beyond
which compatibility with the Planck results for $n_s$ is
jeopardized. These results are also compatible with earlier ones,
obtained, for instance, in Ref.~\cite{Benetti:2016jhf}, when other forms
of an inflaton potential were considered.   An  exception is for the
constant dissipation ratio case, shown by the black curve in
{}Fig.~\ref{fignsXr}. It does allow for fairly large values of
$Q$. {}For instance, for $Q=100$, we have $n_s \simeq 0.96021$ and a
fairly very small $r$ value, $r\simeq 1.1473 \times 10^{-7}$ and it is
still inside the $95\%$ C.L. contour for the Planck 2018 results.

\begin{center}
\begin{figure}[!htb]
\includegraphics[width=8cm]{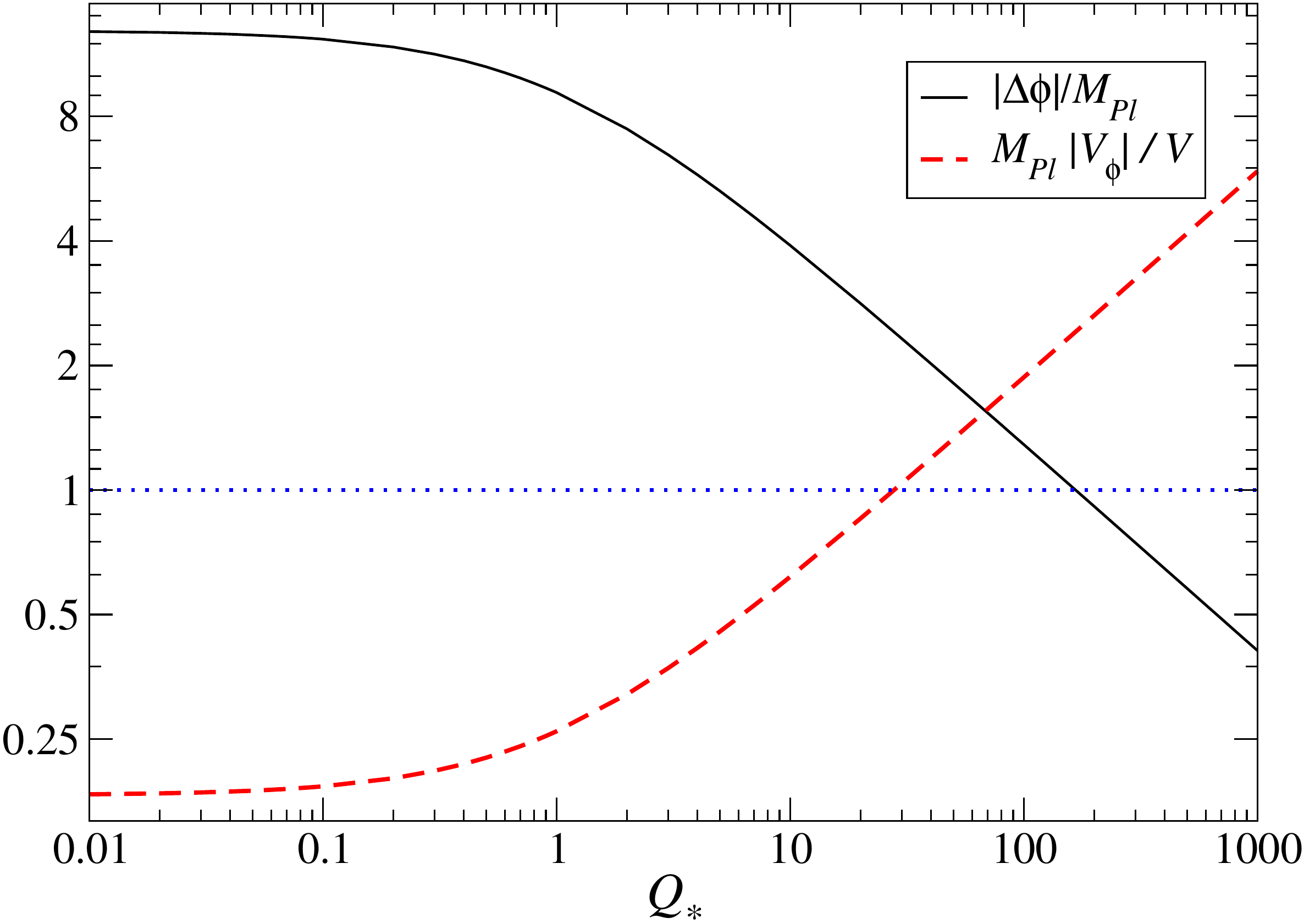}
\caption{The two Swampland conditions Eqs.~(\ref{SDC}) and
  (\ref{SdSC}) as a function of the dissipation ratio $Q_*$.}
\label{fig2}
\end{figure}
\end{center}

As pointed out previously, there are models in WI~\cite{inprogress}
where we can find dissipation coefficients in a high-temperature
regime (with respect to some mass scale, with $T\gg M$) which as the
system evolves goes to a low-temperature ($T < M$) regime. In these
models, it becomes possible to arrange for a dissipation coefficient
that can initially tend to lead to a very red spectrum for large $Q$,
like the inverse in the temperature dissipation coefficient of
Eq.~(\ref{inverse}), but that it ends up transiting at some point to a
dissipation coefficient that tends to lead to a blue spectrum at large
$Q$ values, like the cubic in the temperature form
Eq.~(\ref{cubic}). In these cases it is possible that  $Q$ can be
bigger enough to comply with the bounds set by the Swampland
conjectures.  Even for the dissipation models studied here, if the
Swampland-required values for $\Delta$ and $c$ in Eqs.~(\ref{SDC}) and
(\ref{SdSC}) are slightly relaxed, this can make both conditions be
satisfied already for values of $Q_*$ that are still within the Planck
bounds shown in {}Fig.~\ref{fignsXr}.  In fact, there is quite some
controversy on the precise values for $\Delta$ and $c$, which can also
depend on the specific string model being
considered~\cite{Akrami:2018ylq}. As an example, in {}Fig.~\ref{fig2},
we show the results for the two Swampland conditions as a function of
$Q_*$ when taking the dissipation coefficient case given by
Eq.~(\ref{UpsH}). The other four dissipation coefficients show a
similar evolution for these quantities. We see that, by relaxing the
$\Delta$ value by a factor of around $4$ bigger and decreasing the value
of $c$ by also a factor of $4$ smaller (both with respect to the unit
value), we can already find values of the dissipation ratio, either in the
linear or inverse in $T$ dissipation examples, satisfying both
conditions and still have consistency with the Planck data.

Apart from the above results, one important thing to remark is that WI
breaks the conventional  Lyth bound~\cite{Lyth:1996im}, even for a small
dissipation factor due to the presence of  $
(1+Q)^{-2}\mathcal{F}^{-1}$ in Eq. (\ref{A2}), resulting in a smaller
tensor-to-scalar ratio $r$.  This happens due to the effects of
dissipation and the consequent deviation from the standard
Bunch-Davies vacuum state  (since, in WI, fluctuations are in an excited
state). In fact, a deviation from the standard Bunch-Davies state has
been discussed  as a possible way to save single-field slow-roll
inflation against the Swampland conjectures~\cite{Brahma:2018hrd}.
One may also realize other mechanisms, like the proposal in
Ref.~\cite{Germani:2011bc}, in order to produce a larger $c$ and use
them in the context of WI so as to be able to suppress  the value of $r$,
allowing results to be consistent with observation and still having
small $Q$. {}For instance,  in brane inflation
models~\cite{Bento:2008yx}, we have $\epsilon_{H} =
\epsilon_{V}\frac{4\Lambda}{V}$   with $\Lambda$ here denoting the
brane tension. Then, in the regime where $V\gg \Lambda$, it results in
$r = 24 \epsilon_{H}$. Although the tensor-to-scalar ratio $r$ can
become larger in comparison with  the standard CI,  reconstructing it
in the WI scenario allows one to make it consistent with
observation~\cite{Cid:2007fk} (see also Ref.~\cite{Lin:2018kjm} for a
recent discussion of brane inflation and the Swampland).  Moreover,
all the kinetically modified models of inflation, such as $k$ inflation,
including tachyon and  Dirack-Born-Infeld models, $G$ inflation, and nonminimal derivative
inflation, generally investigated as generalized
$G$ inflation~\cite{Kobayashi:2011nu}, result in an
extra viscous term allowing $c$ to be larger. Although for larger  $c$
they are in general, in the CI picture, in tension with observation,
they can become consistent when constructed in the context of the WI
scenario~\cite{Herrera:2006ckCai:2010wtZhang:2014djaMotaharfar:2017dxhHerrera:2017quxMotaharfar:2018mni}.
Therefore, even if the Swampland conjectures are strictly constrained
to be unity, such an inherent characteristic of WI, i.e., a smaller
energy scale, makes it as a way out of the Swampland.

\section{Conclusions}
\label{concl}

In this work, we have studied the possibility of WI to satisfy the two
recently proposed so-called Swampland conjectures. Despite the fact
that there have been many discussions in the literature about the
validity  of these conjectures and how well they can indeed put
constraints on existing models, particularly those involving inflation
and dark energy, here we have assumed these conjectures to be valid
and discussed WI in that context.

We recall that WI naturally predicts  a lower tensor-to-scalar ratio
$r$ in essentially all models of inflation.  But this becomes
particularly relevant when considering monomial field  chaotic
inflation models, which are assumed already to be ruled out in the CI
scenario. In particular, in the monomial $\phi^4$  model, this feature of
WI in reducing $r$ has already been noticed for almost two
decades~\cite{Berera:1999ws}.  This is of relevance to the Swampland
conjectures, because that implies a lower scale of inflation in WI,
when compared to CI, leading to a natural way for obtaining
sub-Planckian inflaton field values during inflation, even for
monomial chaotic inflation models.  That the energy scale for
inflation in WI can be smaller than that of CI for the same type of
potentials has already been pointed out in
Ref.~\cite{Berera:1998px}. Also, in the review
paper in Ref. ~\cite{BasteroGil:2009ec}, results for the tensor-to-scalar ratio
were explicitly analyzed for monomial models, where it was also
explicitly stated that WI predicts it to be smaller than in CI. This
feature of having a lower $r$ in WI was then explicitly found also in
the more recent
works~\cite{Bastero-Gil:2016qru,Bartrum:2013fiaBerera:2018tfcGraef:2018ulgBastero-Gil:2018uep}
and also when confronting WI with the CMB Planck data through explicit
statistical
analysis~\cite{Benetti:2016jhf,Bastero-Gil:2017wwlArya:2017zlb}.  By
realizing inflation at lower energy scales, then WI makes it possible
to drive different potential models within the safe region based on
the Swampland conjectures. 

{\bf Note added}.--- After completing this paper, a revised
  version of these Swampland conjectures was proposed in
  Ref.~\cite{Ooguri:2018wrx}. Besides the distance conjecture
  Eq.~(\ref{SDC}), in the {\it refined de Sitter conjecture} (RdSC),
  it is required that any scalar field potential from string theory
  should obey either Eq.~(\ref{SdSC}) or $- M^{2}_{\rm Pl}
  (\nabla_{\phi}^2 V)/V > c^{\prime}$, with $c^{\prime}$ being another
  constant also of the order of unity.  This second condition from the
  refined conjecture is related, of course, with the definition of
  the slow-roll coefficient $\eta_V$ in cold inflation.  As we have
  already mentioned in Sec.~\ref{sec2}, in WI  the slow-roll condition
  on $\eta_V$ gets modified by the dissipation ratio $Q$ as $\eta_V <
  1+Q$. Except by the negative sign in the refined conjecture,  which
  restricts the inflaton potential to be of the hilltop, small field
  type, we see that WI for large $Q$ values can still satisfy the
  refined version. In fact, this is related to how WI can provide a
  solution for the so-called $\eta$ problem, as already noticed long
  ago in the very early papers on WI (see, e.g.,
  Ref.~\cite{Berera:1999ws}).

Also, soon after completing this
  paper, another analysis of WI in the context of the Swampland
  conjectures appeared in Ref.~\cite{Das:2018rpg}. The author in that
  reference has reached, however, a different conclusion from us,
  stating that WI was already able to satisfy the conjectures even for
  small values of the dissipation ratio $Q$.  The source of the
  difference can be traced back as a consequence  of mistakenly
  comparing the upper bound coming from  the observation for the
  tensor-to-scale ratio ($r<0.064$) with the lower bound obtained
  from the de Sitter Swampland in
  Ref.~\cite{Das:2018rpg}.  More specifically, our inequality
  Eq.~(\ref{Aq}) is analogous to what is obtained when combining
  Eqs. (19) and (23) of Ref.~\cite{Das:2018rpg}. The difference is
  that the author there takes the function that we have called
  ${\cal F}$ in Eq.~(\ref{Pk}) as $\sim\frac{T}{H}\sqrt{1+Q}$, when
  using an approximated (old) expression for the power spectrum in WI.
  Note that when considering the correct expression that we have used
  here, given by Eq.~(\ref{Pk}), and for the case of temperature-independent dissipation, we can then set $G(Q)=1$ in
  Eq.~(\ref{calF}). Thus, for the case of $Q \ll 1$, we have ${\cal F}
  \sim 2 T/H$ (for $T/H \gg 1$, when still in the WI regime), and then our
  Eq.~(\ref{Aq})  becomes $4 c^2 (H/T) < r < 4 \Delta^2 (1/N^{2})
  (H/T)$. It then becomes obvious that it is quite hard for the
  inequalities to be {\it simultaneously } satisfied unless we relax
  severely either on $c$, on $\Delta$, or on both. However, for $Q
  \gg 1$, then ${\cal F} \sim \sqrt{3 \pi Q} T/H$ and now
  Eq.~(\ref{Aq})  becomes $(8c^2 /\sqrt{3 \pi})(H/T)(1/Q)^{5/2} < r <
  (8\Delta^2/\sqrt{3 \pi Q}) (1/N^{2}) (H/T)$.  We can now understand why
  for large $Q$ both inequalities can be satisfied.  This happens
  provided that $c^2/Q^2 < \Delta^2/N^{2}$, which is actually just
  another way of writing Eq.~(\ref{A1}). {}Furthermore, the results
  present here in {}Fig.~\ref{fig2} illustrate quite well the same
  result, independent of the derivation leading to Eq.~(\ref{A1}) or
  Eq.~(\ref{Aq}).  Moreover, it is  worth mentioning that a
  different analysis performed recently by the authors of
  Ref.~\cite{Bastero-Gil:2018yen} has also reached a similar
  conclusion to ours.

\section*{Acknowledgments}

R.O.R. is partially supported by research grants from Conselho Nacional
de Desenvolvimento Cient\'{\i}fico e Tecnol\'ogico (CNPq), Grant
No. 302545/2017-4, and Funda\c{c}\~ao Carlos Chagas Filho de Amparo \`a
Pesquisa do Estado do Rio de Janeiro (FAPERJ), Grant No. E-26/202.892/2017. 



\begin{thebibliography}{99}

\bibitem{Aghanim:2018eyx}  N.~Aghanim {\it et al.} (Planck
  Collaboration), { \it Planck 2018 results. VI. Cosmological
  parameters}, arXiv:1807.06209.  
  


\bibitem{inflation} A.~A.~Starobinsky, {\it A new type of isotropic
  cosmological models without singularity}, Phys.\ Lett.\ {\bf 91B},
  99 (1980); K.~Sato, {\it First order phase transition of a vacuum
    and expansion of the universe},
  Mon.\ Not.\ R.\ Astron.\ Soc.\  {\bf 195}, 467 (1981); A.~H.~Guth,
  {\it The inflationary Universe: A possible solution to the horizon
    and flatness problems}, Phys.\ Rev.\ D {\bf 23}, 347 (1981);
  A.~Albrecht and P.~J.~Steinhardt, {\it Cosmology for grand unified
    theories with radiatively induced symmetry breaking},
  Phys.\ Rev.\ Lett.\  {\bf 48}, 1220 (1982); A.~D.~Linde, {\it A new
    inflationary Universe scenario: A possible solution of the
    horizon, flatness, homogeneity, isotropy and primordial monopole
    problems}, Phys.\ Lett.\ {\bf 108B}, 389 (1982). 


  
\bibitem{Vafa}
C. Vafa, {\it The string landscape and the Swampland}, arXiv:hep-th/0509212.

\bibitem{Ooguri:2006in}
  H.~Ooguri and C.~Vafa,
 {\it On the geometry of the string landscape and the Swampland}, Nucl.\ Phys.\ {\bf B766}, 21 (2007),
  arXiv:hep-th/0605264.


\bibitem{Palti:2017elp} 
  E.~Palti,
  {\it The weak gravity conjecture and scalar fields}, J. High Energy Phys. {08} (2017) 034,
  arXiv:1705.04328.
  

\bibitem{Obied:2018sgi}
  G.~Obied, H.~Ooguri, L.~Spodyneiko and C.~Vafa, {\it De Sitter space and the Swampland},
  arXiv:1806.08362.

\bibitem{Arkani-Hamed}
N. Arkani-Hamed, L. Motl, A. Nicolis, and C. Vafa,
{\it The string landscape, black holes and gravity as the weakest force} J. High Energy Phys. {06} (2007) 060, arXiv:hep-th/0601001.

\bibitem{Ooguri:qwFreivoge:swBrennan:sd}
H. Ooguri and C. Vafa, {\it Non-supersymmetric AdS and the Swampland}, Adv. Theor. Math. Phys. {\bf 21},
1787 (2017), arXiv:1610.01533;
B. Freivogel and M. Kleban, {\it Vacua morghulis}, arXiv:1610.04564;
T. D. Brennan, F. Carta, and C. Vafa, {\it The string landscape, the Swampland, and the missing corner}, Proc. Sci. TASI2017 (2017) 015, arXiv:1711.00864.

\bibitem{Agrawal:2018own} 
  P.~Agrawal, G.~Obied, P.~J.~Steinhardt, and C.~Vafa,
  ``On the cosmological implications of the string Swampland,''
  Phys.\ Lett.\ B {\bf 784}, 271 (2018), arXiv:1806.09718.

\bibitem{Dvali:dc}
G. Dvali and C. Gomez, { \it On exclusion of positive cosmological constant}, Fortschr. Phys.
{\bf 67}, 1800092 (2019), arXiv:1806.10877.

\bibitem{Andriot:2018eptRoupec:2018mbnAndriot:2018wzkColgain:2018wgkHeisenberg:2018yae} 
  D.~Andriot, {\it New constraints on classical de Sitter: Flirting with the Swampland}, Fortschr. Phys. {\bf 67}, 1800103 (2019), arXiv:1807.09698;
  C.~Roupec and T.~Wrase, {\it de Sitter extrema and the Swampland }, Fortschr. Phys. {\bf 67}, 1800082 (2019),
  arXiv:1807.09538;
  D.~Andriot, {\it On the de Sitter Swampland criterion}, Phys. Lett. B {\bf 785}, 570 (2018), arXiv:1806.10999;
  E.~O.~Colgain, M.~H.~P.~M.~Van Putten, and H.~Yavartanoo, {\it $H_0$ tension and the de Sitter Swampland},
  arXiv:1807.07451;
  L.~Heisenberg, M.~Bartelmann, R.~Brandenberger, and A.~Refregier, {\it Dark energy in the Swampland }, Phys.Rev. D {\bf 98}, 123502 (2018), arXiv:1808.02877.


\bibitem{Garg:2018reuBrown:2015ihaMatsui:2018bsyBen-Dayan:2018mheKinney:2018nny} 
  S.~K.~Garg and C.~Krishnan, {\it Bounds on slow roll and the de Sitter Swampland},
  arXiv:1807.05193;
  J.~Brown, W.~Cottrell, G.~Shiu, and P.~Soler,
  {\it Fencing in the Swampland: Quantum gravity constraints on
                        large field inflation}, J. High Energy Phys. {10} (2015) 023, arXiv:1503.04783;
  H.~Matsui and F.~Takahashi, {\it Eternal inflation and Swampland conjectures}, Phys.Rev. D {\bf 99}, 023533 (2019), arXiv:1807.11938;
  I.~Ben-Dayan,
  {\it Draining the Swampland},
  arXiv:1808.01615;
  W.~H.~Kinney, S.~Vagnozzi, and L.~Visinelli,
  {\it The zoo plot meets the Swampland: Mutual (in)consistency of single-field inflation, string conjectures, and cosmological data},
  arXiv:1808.06424.
  
  
\bibitem{Dias:2018ngv} 
  M.~Dias, J.~Frazer, A.~Retolaza, and A.~Westphal, {\it Primordial gravitational waves and the Swampland}, Fortschr. Phys. {\bf 67}, 1800063 (2019),
  arXiv:1807.06579.



\bibitem{Agrawal:2018}
P.~ Agrawal, J.~ Fan, and M.~ Reece, {\it Clockwork axions in cosmology: is chromonatural inflation chrononatural?}, J. High Energy Phys. {10} (2018) 193, arXiv:1806.09621.


\bibitem{Achucarro:2018vey} 
  A.~Achúcarro and G.~A.~Palma, {\it The string Swampland constraints require multi-field inflation }, J. Cosmol. Astropart. phys. {02} (2019) 041, arXiv:1807.04390.

\bibitem{Kehagias:2018uem} 
  A.~Kehagias and A.~Riotto, {\it A note on inflation and the Swampland}, Fortschr. Phys. {\bf 66}, 1800052  (2018), arXiv:1807.05445.
  
\bibitem{Akrami:2018ylq} 
  Y.~Akrami, R.~Kallosh, A.~Linde, and V.~Vardanyan,
  {\it The landscape, the Swampland and the era of precision cosmology}, Fortschr. Phys. {\bf 67}, 1800075 (2019), arXiv:1808.09440.
  
\bibitem{Conlon:2018eyr} 
  J.~P.~Conlon,
  {\it The de Sitter Swampland conjecture and supersymmetric AdS vacua}, Int. J. Mod. Phys. A {\bf 33}, 1850178 (2018), arXiv:1808.05040.
  
  
\bibitem{Berera:1995ie} 
  A.~Berera,
  {\it Warm Inflation},
   Phys.\ Rev.\ Lett.\  {\bf 75}, 3218 (1995), arXiv:astro-ph/9509049.
  
\bibitem{Berera:2008ar} 
  A.~Berera, I.~G.~Moss, and R.~O.~Ramos,
  {\it Warm inflation and its microphysical basis},
   Rep.\ Prog.\ Phys.\  {\bf 72}, 026901 (2009), arXiv:0808.1855.
  
\bibitem{BasteroGil:2009ec} 
  M.~Bastero-Gil and A.~Berera,
  {\it Warm inflation model building},
   Int.\ J.\ Mod.\ Phys.\ A {\bf 24}, 2207 (2009), arXiv:0902.0521.
  
  
\bibitem{Das:2018hqy} 
  S.~Das,
  {\it A note on single-field inflation and the Swampland criteria},
  arXiv:1809.03962.

\bibitem{ABCM}
R.~Allahverdi, R.~Brandenberger, F.~Y.~Cyr-Racine, and A.~Mazumdar,
  {\it Reheating in inflationary cosmology: Theory and applications},
  Annu.\ Rev.\ Nucl.\ Part.\ Sci.\  {\bf 60}, 27 (2010), arXiv:1001.2600.
  
\bibitem{Karouby}  
M.~A.~Amin, M.~P.~Hertzberg, D.~I.~Kaiser, and J.~Karouby,
  {\it Nonperturbative dynamics of reheating after inflation: A review},
  Int.\ J.\ Mod.\ Phys.\ D {\bf 24}, 1530003 (2014), arXiv:1410.3808.

  
\bibitem{BasteroGil:2010pb} 
  M.~Bastero-Gil, A.~Berera, and R.~O.~Ramos,
  {\it Dissipation coefficients from scalar and fermion quantum field interactions},
  J. Cosmol. Astropart. Phys. {09} (2011) 033, arXiv:1008.1929.

\bibitem{BasteroGil:2012cm} 
  M.~Bastero-Gil, A.~Berera, R.~O.~Ramos, and J.~G.~Rosa,
  {\it General dissipation coefficient in low-temperature warm inflation},
  J. Cosmol. Astropart. Phys. {01} (2013) 016, arXiv:1207.0445.
    
    
\bibitem{Hall:2003zp}
  L.~M.~H.~Hall, I.~G.~Moss, and A.~Berera,
\emph{Scalar perturbation spectra from warm inflation},
  Phys.\ Rev.\ D {\bf 69},  083525 (2004), arXiv:astro-ph/0305015.
  
\bibitem{Graham:2009bf} 
  C.~Graham and I.~G.~Moss,
\emph{Density fluctuations from warm inflation},
  J. Cosmol. Astropart. Phys. {07} (2009) 013, arXiv:0905.3500.


\bibitem{BasteroGil:2011xd} 
  M.~Bastero-Gil, A.~Berera, and R.~O.~Ramos,
\emph{Shear viscous effects on the primordial power spectrum from warm inflation},
  J. Cosmol. Astropart. Phys. {07} (2011) 030, arXiv:1106.0701.
 
\bibitem{Bastero-Gil:2014jsa} 
  M.~Bastero-Gil, A.~Berera, I.~G.~Moss, and R.~O.~Ramos,
  {\it Cosmological fluctuations of a random field and radiation fluid},
   J. Cosmol. Astropart. Phys. {05} (2014) 004, arXiv:1401.1149.
  

\bibitem{Bastero-Gil:2014raa} 
  M.~Bastero-Gil, A.~Berera, I.~G.~Moss, and R.~O.~Ramos,
\emph{Theory of non-Gaussianity in warm inflation},
  J. Cosmol. Astropart. Phys. {12} (2014) 008, arXiv:1408.4391.


\bibitem{Visinelli:2014qla} 
  L.~Visinelli,
  {\it Cosmological perturbations for an inflaton field coupled to radiation},
  J. Cosmol. Astropart. Phys. {\bf 01} (2015) 005, arXiv:1410.1187.


\bibitem{Benetti:2016jhf} 
  M.~Benetti and R.~O.~Ramos,
  {\it Warm inflation dissipative effects: Predictions and constraints from the Planck data},
  Phys.\ Rev.\ D {\bf 95}, 023517 (2017), arXiv:1610.08758.


\bibitem{Ramos:2013nsa} 
  R.~O.~Ramos and L.~A.~da Silva,
  {\it Power spectrum for inflation models with quantum and thermal noises},
   J. Cosmol. Astropart. Phys. {03} (2013) 032, arXiv:1302.3544.
  
  
  
\bibitem{Li:2018wno} 
  X.~B.~Li, H.~Wang, and J.~Y.~Zhu,
{\it Gravitational waves from warm inflation},
  Phys.\ Rev.\ D {\bf 97}, 063516 (2018), arXiv:1803.10074.
  
  
\bibitem{Bastero-Gil:2016qru} 
  M.~Bastero-Gil, A.~Berera, R.~O.~Ramos, and J.~G.~Rosa,
  {\it Warm Little Inflaton},
  Phys.\ Rev.\ Lett.\  {\bf 117}, 151301 (2016), arXiv:1604.08838.
   

\bibitem{Berera:1998gx}
  A.~Berera, M.~Gleiser, and R.~O.~Ramos,
\emph{Strong dissipative behavior in quantum field theory},
  Phys.\ Rev.\ D {\bf 58},  123508 (1998), arXiv:hep-ph/9803394.


\bibitem{Berera:1998px}
  A.~Berera, M.~Gleiser, and R.~O.~Ramos,
\emph{A First principles warm inflation model that solves the cosmological horizon / flatness problems},
  Phys.\ Rev.\ Lett.\  {\bf 83},  264 (1999), arXiv:hep-ph/9809583.


\bibitem{Zhang:2009geHerrera:2014mcaJawad:2017gwa} 
  Y.~Zhang,
  {\it Warm inflation with a general form of the dissipative coefficient},
  J. Cosmol. Astropart. Phys. {03} (2009) 023, arXiv:0903.0685;
  R.~Herrera, M.~Olivares, and N.~Videla,
  {\it General dissipative coefficient in warm intermediate inflation in loop quantum cosmology in light of Planck and BICEP2},
  Int.\ J.\ Mod.\ Phys.\ D {\bf 23}, 1450080 (2014), arXiv:1404.2803;
  A.~Jawad, S.~Hussain, S.~Rani, and N.~Videla,
  {\it Impact of generalized dissipative coefficient on warm inflationary dynamics in the light of latest Planck data},
  Eur.\ Phys.\ J.\ C {\bf 77}, 700 (2017), arXiv:1709.10430.
  
  
\bibitem{Bolotin:2013jpa}
  Y.~L.~Bolotin, A.~Kostenko, O.~A.~Lemets, and D.~A.~Yerokhin,
  {\it Cosmological evolution with interaction between dark energy and dark matter},
  Int.\ J.\ Mod.\ Phys.\ D {\bf 24}, 1530007 (2014), arXiv:1310.0085.


\bibitem{Barbosa:2017ojt} 
  C.~M.~S.~Barbosa, H.~Velten, J.~C.~Fabris, and R.~O.~Ramos,
  {\it Assessing the impact of bulk and shear viscosities on large scale structure formation},
  Phys.\ Rev.\ D {\bf 96}, 023527 (2017),
  arXiv:1702.07040.
  


\bibitem{Akrami:2018odb} 
  Y.~Akrami {\it et al.}, (Planck Collaboration), {\it Planck 2018 results. X. Constraints on inflation},
  arXiv:1807.06211.


\bibitem{inprogress} M.~Bastero-Gil, A.~Berera, R.~O.~Ramos, and J.~G.~Rosa
  {\it (work in progress)}.
  
  
\bibitem{Lyth:1996im} 
  D.~H.~Lyth,
  {\it What would we learn by detecting a gravitational wave signal in the cosmic microwave background anisotropy?},
  Phys.\ Rev.\ Lett.\  {\bf 78}, 1861 (1997), arXiv:hep-ph/9606387.

\bibitem{Brahma:2018hrd}
  S.~Brahma and M.~Wali Hossain,
 {\it Avoiding the string Swampland in single-field inflation: Excited initial states},
  arXiv:1809.01277.

\bibitem{Germani:2011bc}
  C.~Germani, L.~Martucci, and P.~Moyassari,
 {\it Introducing the Slotheon: A slow Galileon scalar field in curved space-time},
  Phys.\ Rev.\ D {\bf 85} 103501 (2012),
  arXiv:1108.1406.
  
\bibitem{Bento:2008yx}
  M.~C.~Bento, R.~G.~Felipe, and N.~M.~C.~Santos,
{\it Brane assisted quintessential inflation with transient acceleration},
  Phys.\ Rev.\ D {\bf 77}, 123512 (2008), arXiv:0801.3450.

\bibitem{Cid:2007fk}
  M.~A.~Cid, S.~del Campo, and R.~Herrera,
  {\it Warm inflation on the brane},
  J. Cosmol. Astropart. Phys. {10} (2007) 005,
  arXiv:0710.3148.
  
  
\bibitem{Lin:2018kjm} 
  C.~M.~Lin, K.~W.~Ng, and K.~Cheung,
  {\it Chaotic inflation on the brane and the Swampland criteria},
  arXiv:1810.01644.
  

\bibitem{Kobayashi:2011nu}
  T.~Kobayashi, M.~Yamaguchi, and J.~Yokoyama,
 {\it Generalized $G$ inflation: Inflation with the most general second-order field equations},
  Prog.\ Theor.\ Phys.\  {\bf 126}, 511 (2011), arXiv:1105.5723.

\bibitem{Herrera:2006ckCai:2010wtZhang:2014djaMotaharfar:2017dxhHerrera:2017quxMotaharfar:2018mni}
  R.~Herrera, S.~del Campo, and C.~Campuzano,
 {\it Tachyon warm inflationary universe models},
  J. Cosmol. Astropart. Phys. {\bf 10} (2006) 009, arXiv:astro-ph/0610339;
  Y.~F.~Cai, J.~B.~Dent, and D.~A.~Easson,
 {\it Warm DBI inflation},
  Phys.\ Rev.\ D {\bf 83}, 101301 (2011), arXiv:1011.4074;
  X.~M.~Zhang and j.~Y.~Zhu,
 {\it Extension of warm inflation to noncanonical scalar fields},
  Phys.\ Rev.\ D {\bf 90}, 123519 (2014), arXiv:1402.0205.
  M.~Motaharfar, E.~Massaeli, and H.~R.~Sepangi,
  {\it Power spectra in warm $G$ inflation and its consistency: Stochastic approach},
  Phys.\ Rev.\ D {\bf 96},  103541 (2017), arXiv:1705.04049;
  R.~Herrera,
  {\it G-Warm inflation},
  J. Cosmol. Astropart. Phys. {05} (2017) 029, arXiv:1701.07934;
  M.~Motaharfar, E.~Massaeli, and H.~R.~Sepangi,
  {\it Warm Higgs $G$ inflation: Predictions and constraints from Planck 2015 likelihood}, J. Cosmol. Astropart. Phys. {10} (2018) 002, arXiv:1807.09548.

  
  
\bibitem{Berera:1999ws} A.~Berera, 
{\it Warm inflation at arbitrary adiabaticity: A Model, an
existence proof for inflationary dynamics in quantum field
theory},  
Nucl.\ Phys.\ {\bf B585}, 666 (2000), arXiv:hep-ph/9904409.
  



\bibitem{Bartrum:2013fiaBerera:2018tfcGraef:2018ulgBastero-Gil:2018uep} 
  S.~Bartrum, M.~Bastero-Gil, A.~Berera, R.~Cerezo, R.~O.~Ramos, and J.~G.~Rosa,
  {\it The importance of being warm (during inflation)},
   Phys.\ Lett.\ B {\bf 732}, 116 (2014), arXiv:1307.5868;
  A.~Berera, J.~Mabillard, M.~Pieroni, and R.~O.~Ramos,
  {\it Identifying universality in warm inflation},
  J. Cosmol. Astropart. Phys. {07} (2018) 021, arXiv:1803.04982;
  L.~L.~Graef and R.~O.~Ramos,
  {\it Probability of warm inflation in loop quantum cosmology},
  Phys.\ Rev.\ D {\bf 98}, 023531 (2018), arXiv:1805.05985;
  M.~Bastero-Gil, A.~Berera, R.~Hernandez-Jimenez, and J.~G.~Rosa,
  {\it Dynamical and observational constraints on the Warm Little Inflaton scenario}, arXiv:1805.07186.
  
\bibitem{Bastero-Gil:2017wwlArya:2017zlb} 
  M.~Bastero-Gil, S.~Bhattacharya, K.~Dutta, and M.~R.~Gangopadhyay,
{\it Constraining warm inflation with CMB data},
  J. Cosmol. Astropart. Phys. {02} (2018) 054, arXiv:1710.10008;
  R.~Arya, A.~Dasgupta, G.~Goswami, J.~Prasad, and R.~Rangarajan,
{\it Revisiting CMB constraints on warm inflation},
  J. Cosmol. Astropart. Phys. {02} (2018) 043, arXiv:1710.11109.
  
\bibitem{Ooguri:2018wrx} 
  H.~Ooguri, E.~Palti, G.~Shiu, and C.~Vafa,
  {\it Distance and de Sitter conjectures on the Swampland},
  Phys.\ Lett.\ B {\bf 788}, 180 (2019), arXiv:1810.05506.

 
\bibitem{Das:2018rpg} 
  S.~Das,
  {\it Warm inflation in the light of Swampland criteria}, 
  arXiv:1810.05038.

\bibitem{Bastero-Gil:2018yen} 
  M.~Bastero-Gil, A.~Berera, R.~Hern\'andez-Jim\'enez, and J.~G.~Rosa,
  {\it Warm inflation within a supersymmetric distributed mass model},
  arXiv:1812.07296.

\end{thebibliography}
\end{document}